\documentclass[12pt,a4paper]{article}
\usepackage{amsmath,amssymb}

\newcommand{\be}{\begin{eqnarray}}
\newcommand{\ee}{\end{eqnarray}}

\begin{document}

\title{\textbf{Similarity Solutions of a Class of Perturbative Fokker-Planck Equation}}

\author{Wen-Tsan Lin and Choon-Lin Ho\\
Department of Physics, Tamkang University\\
Tamsui 25137, Taiwan, R.O.C.}



\maketitle

\begin{abstract}

In a previous work, a perturbative approach to a class of
Fokker-Planck equations, which have constant diffusion
coefficients and small time-dependent drift coefficients, was
developed by exploiting the close connection between the
Fokker-Planck equations and the Schr\"odinger equations.  In this
work, we further explore the possibility of similarity solutions
of such a class of Fokker-Planck equations. These solutions possess
definite scaling behaviors, and are obtained by means of the
so-called similarity method.

\end{abstract}



\newpage

\section{Introduction}

One of the basic equations employed to describe fluctuating
macroscopic variables is the Fokker-Planck equation (FPE)
\cite{RIS:1996}. It has found applications not only in physics,
but also in other areas such as astrophysics
\cite{NM:1998,HCDHQS:2006}, chemistry \cite{NOW:1996,LHE:1995},
biology \cite{SG:1993,FBF:2003}, finance \cite{FPR:2000}, etc. In
view of its wide applicability, various methods of finding exact
and approximate solutions of the FPE's have been developed
\cite{RIS:1996,BS:1985,MP:1996,BM:1996,HU:1989,LLPS:1985,LAN:1991}.
Most of the methods, however, are concerned only with FPE's with
time-independent diffusion and drift coefficients. Generally, it
is not easy to find solutions of FPE's with time-dependent
diffusion and drift coefficients.

One of the methods of solving FPE with time-independent diffusion
and drift coefficients is to transform the FPE into a
time-independent Schr\"odinger equation, and then solve the
eigenvalue problem of the latter \cite{RIS:1996,HS:2008}. The
transformation to the Schr\"odinger equation of a FPE eliminates
the first order spatial derivative in the FPE and creates a
Hermitian spatial differential operator.  In [16] it was shown
that this transformation can also be applied to transform FPE's,
which have constant diffusion coefficients and time-dependent
drift coefficients, into time-dependent Schr\"odinger equations.
Based on this connection, a perturbative approach was developed
and used to solve this special class of the FPE \cite{HD:2008}.

The diffusion equation is known to be invariant under the scale
transformation $x\rightarrow\varepsilon x$, $t\rightarrow\varepsilon^2 t$ for any
scale $\varepsilon$.  Thus one can find solutions, called the
similarity solutions, of the diffusion equation with definite
scaling behaviors by means of the so-called similarity method
\cite{ZIL:1992}. Such solutions were not considered in [16]. It is
the purpose of this work to investigate the possibility of
similarity solutions of the FPE within the perturbative framework
of [16].

\section{Perturbative approach to Fokker-Planck equations}

The FPE of the probability density $W(x,t)$ in $(1+1)$-dimension
is \cite{RIS:1996}
\begin{align}\label{PA1.1}
\frac{\partial}{\partial t}W(x,t)=\Big[-\frac{\partial}{\partial x}D^{(1)}(x,t)+\frac{\partial^2}
{\partial x^2}D^{(2)}(x,t)\Big]\,W(x,t)\;,
\end{align}
where $D^{(1)}(x,t)$ and $D^{(2)}(x,t)$ are the drift and
diffusion coefficient respectively. The drift coefficient
represents the external force acting on the particles and is
usually expressed in terms of a drift potential $U(x,t)$ according
to $D^{(1)}(x,t)=-\partial U(x,t)/\partial x$.

In this paper we will focus on FPE's with constant diffusion
coefficients $D^{(2)}(x,t)=D>0$.  In this case, one can solve the
FPE by exploiting the connection between the FPE and the
Schr\"odinger equation \cite{RIS:1996,HS:2008}.  Setting
\begin{align}\label{PA1.2}
\psi(x,t)=\mbox{exp}\Big\{\frac{U(x,t)}{2D}\Big\}\,W(x,t),
\end{align}
one can transfrom the FPE into the Schr\"odinger equation
\begin{align}\label{PA1.3}
\frac{\partial \psi}{\partial t}=D\frac{\partial^2 \psi}{\partial x^2}
+\Big(\frac{U''}{2}-\frac{U'^2}{4D}
+\frac{\dot{U}}{2D}\Big)\psi\;,
\end{align}
where the prime and dot denote the derivatives with respect to $x$
and $t$, respectively. Each solution of the time-dependent
Schr\"odinger equation gives the corresponding solution of the FPE
via Eq.~(\ref{PA1.2}). It is, however, generally difficult to
solve the Schr\"odinger equation exactly when the potential is
time-dependent.  Various approximation schemes may have to be
employed.

The perturbative approach presented in [16] to solve the FPE
containing a small parameter in the drift potential is summarized
as follows. Suppose $U(x,t)=\sum^{\infty}_{n=0}\lambda^nU_n(x,t)$,
where $|\lambda|\ll 1$ is a small parameter, and let
$\psi(x,t)=\mbox{exp}\{S(x,t,\lambda)/D\}$. Then from
(\ref{PA1.3}), we get the equation of $S(x,t,\lambda)$,
\begin{align}
\dot{S}&=DS''+S'^2+\overline{U}\;,\label{PA1.5}\\
\overline{U}&\equiv\frac{D}{2}U''-\frac{1}{4}U'^2
+\frac{1}{2}\dot{U}\;.\label{PA1.6}
\end{align}
Substituting the series form of
$S(x,t,\lambda)=\sum^{\infty}_{n=0}\,\lambda^n S_n(x,t)$ into
(\ref{PA1.5}) and collecting terms of the same order of $\lambda$,
one arrives at a set of differential equations that determine the
functions $S_n$:
\begin{align}
\dot{S}_0&=DS_0''+S_0'^2+\Big(\frac{D}{2}U_0''-\frac{1}{4}U_0'^2
+\frac{1}{2}\dot{U}_0\Big)\;,\label{PA1.7}\\
\dot{S}_1&=DS_1''+2S_0'S_1'+\Big(\frac{D}{2}U_1''-\frac{1}{2}U_0'\,
U_1'+\frac{1}{2}\dot{U}_1\Big)\;,\label{PA1.8}\\
\dot{S}_2&=DS_2''+2S_0'S_2'+S_1'^2+\Big(\frac{D}{2}U_2''
-\frac{1}{2}U_0'U_2'-\frac{1}{4}U_1'^2
+\frac{1}{2}\dot{U}_2\Big)\;,\label{PA1.9}\\
\vdots\notag\\
\dot{S}_n&=DS_n''+\sum^{\infty}_{k=0}S_k'S'_{n-k}+\mbox{terms in
$\overline{U}$ of the order $\lambda^n$}\;,\;\;\;n\geq
0\;.\label{PA1.10}
\end{align}
With $S_0$, $S_1$, $S_2$,... solved, we will have an approximate
solution $\psi(x,t)$, and hence of $W(x,t)$.

In this paper, we shall be interested in the drift potential
$U(x,t)$ of the form $U(x,t)=\lambda\,U_1(x,t)$, with all other
$U_n=0$ for $n=0$ and $n>1$. From (\ref{PA1.2}) the probability
density is
\begin{align}\label{PA1.11}
W(x,t)=e^{\frac{1}{D}(S-\frac{U}{2})}=e^{\frac{S_0}{D}}\,e^{\frac{1}{D}(\,\sum^{\infty}_{n=1}
\lambda^nS_n-\lambda\frac{U_1}{2})}\;,
\end{align}
We assume the initial profile of $W(x,t)$ of the unperturbed case
(the diffusion case) to be the delta-function, i.e.,
$W(x,t)\rightarrow\delta(x)$ at $t=0$ as $\lambda\rightarrow 0$.
Then the solution $S_0(x,t)$ is \cite{HD:2008}
\begin{align}\label{PA1.12}
S_0(x,t)=-\frac{D}{2}\ln(4\pi Dt)-\frac{x^2}{4t}\;,
\end{align}
leading to the probability density
\begin{align}\label{PA1.13}
W_0(x,t)=e^{\frac{S_0}{D}}=\frac{1}{\sqrt{4\pi
Dt}}\,e^{-\frac{x^2}{4Dt}}\;.
\end{align}
Eq.~(\ref{PA1.13}) is the well-known solution of the diffusion
equation with the delta function as the initial profile. It is
evident that under the scale transformation
$x\rightarrow\bar{x}=\varepsilon x$,
$t\rightarrow\bar{t}=\varepsilon^2 t$, $W_0(x,t)$ scales as
$W_0(x,t)=\varepsilon W_0(\bar{x},\bar{t})$.

\section{Scaling behavior}

Motivated by the scaling form of $W_0(x,t)$ in (\ref{PA1.13}), we
would like to seek similarity solutions of the perturbative FPE.
To this end, let us first obtain the required scaling behaviors of
$W(x,t)$ and $U(x,t)$ so that Eq.~(\ref{PA1.1}) is invariant under
the scale transformation
\begin{equation}
\bar{x}=\varepsilon^a x,~~\bar{t}=\varepsilon^b t, \label{scale1}
\end{equation}
where the scaling exponents $a$ and $b$ are arbitrary real
parameters. Suppose $W(x,t)$ and $U(x,t)$ scale as
$\bar{W}(\bar{x},\bar{t})=\varepsilon^{\gamma}W(x,t)$ and
$\bar{U}(\bar{x},\bar{t})=\varepsilon^{d}U(x,t)$ with real
exponent $\gamma$ and $d$, resepctively. Then Eq.~(\ref{PA1.1})
becomes
\begin{align}
\label{sb.1}
\varepsilon^{-\gamma+b}\frac{\partial
\bar{W}}{\partial
\bar{t}}=\varepsilon^{-\gamma-d+2a}\frac{\partial } {\partial
\bar{x}}\Big(\frac{\partial \bar{U}}{\partial \bar{x}}\cdot
\bar{W}\Big)+\varepsilon^{-\gamma+2a}
D\frac{\partial^2\bar{W}}{\partial \bar{x}^2}\;.
\end{align}
For Eq.(\ref{sb.1}) to have the same form as Eq.~(\ref{PA1.1}),
one must have $\varepsilon^{-\gamma+b} =\varepsilon^{-\gamma-d+2a}
=\varepsilon^{-\gamma+2a}$.  This implies that $d=0$, $b=2a$, and
$\gamma$ arbitrary. It means that when the scaling exponent of $t$
is twice that of $x$, and $U(x,t)$ is scale-invariant, then the
FPE is scale invariant, and thus admits solutions $W(x,t)$ with
arbitrary scaling exponent, which is dictated by the initial
profile $W(x,0)$. This result puts a constraint on the scaling
property of the drift potential $U(x,t)$.

Applying these general results to Eq.~(\ref{PA1.11}), we see that
since $U(x,t)=\lambda U_1(x,t)$ is scale-invariant, so are all
$S_n (x,t)~(n\geq 1)$. The scaling of $W$ is therefore solely
determined by the solution $\exp\{S_0/D\}$ of the non-perturbed
FPE.  With $S_0(x,t)$ given in (\ref{PA1.12}), $W(x,t)$ scales as
$W(x,t)=\varepsilon^a W(\bar{x},\bar{t})$ (see Eq.(\ref{PA1.13})).
This is in accord with the scaling of the initial profile:
$W(x,0)=\delta(x)=\varepsilon^a \delta(\bar{x})$. Our next step is
to obtain the scale-invariant solutions for the $S_n$'s.  This is
attained by the similarity method, which we describe below.

\section{Similarity method}

The similarity method is a very useful method for solving a
partial differential equation which possesses proper scaling
behavior. One advantage of the similarity method is to reduce the
partial differential equation to an ordinary differential equation
through some new independent variables (called similarity
variables), which are certain combinations of the old independent
variables. In our case, the 2nd order FPE can thus be transformed
into an ordinary differential equation which may be easier to
solve.  We will illustrate the method by discussing the solution
of $S_1$ below. The method applies to equations for other $S_n$ as
well.

With $U_0(x,t)=0$ and $S_0$ given in (\ref{PA1.12}), the equation
for $S_1$ is
\begin{align}\label{E1.2}
\dot{S}_1=DS_1''-\frac{x}{t}S'_1+\frac{D}{2}U_1''+\frac{1}{2}\dot{U}_1\;.
\end{align}
Let $x$ and $t$ transform according to Eq.~(\ref{scale1}). Recall
from previous discussions that $U_1$ and $S_n$ ($n\geq 1$) are
scale-invariant, i.e., $U(x,t)=\bar{U}_1(\bar{x},\bar{t})$ and
$S_n(x,t)=\bar{S}_n(\bar{x},\bar{t})$ ($n\geq 1$).  In terms of
these scaled variables, Eq.(\ref{E1.2}) becomes
\begin{align}
\varepsilon^{b}\frac{\partial \bar{S}_1}{\partial
\bar{t}}&=\varepsilon^{2a}D \frac{\partial^2 \bar{S}_1}{\partial
\bar{x}^2}-\varepsilon^{b}\frac{\bar{x}}{\bar{t}} \frac{\partial
\bar{S}_1}{\partial
\bar{x}}+\varepsilon^{2a}\frac{D}{2}\frac{\partial^2
\bar{U}_1}{\partial
\bar{x}^2}+\varepsilon^{b}\frac{1}{2}\frac{\partial
\bar{U}_1}{\partial \bar{t}}\;.\label{E1.7}
\end{align}
We demand that $\bar{S}_1$ satisfies Eq.(\ref{E1.2}) and this
requires $b=2a$. To determine the form of the similarity solution,
we have to first determine a new variable $z(x,t)$ that is
invariant under the scale transformation Eq.(\ref{scale1}). From
$b=2a$, we can choose the scale-invariant similarity variable to
be
\begin{align}\label{E1.11}
z=\frac{x}{\sqrt{t}}\;.
\end{align}

Since both $S_1(x,t)$ and $U_1(x,t)$ are scaling invariant, they
must be functions of $z$ only. Hence it is reasonable to express
$S_1(x,t)$ and $U_1(x,t)$ in terms of similarity variable $z$ as
\begin{align}\label{E1.15}
S_1(x,t)=y(z);\;\;\;U_1(x,t)=u(z)\;.
\end{align}
Eq.(\ref{E1.2}) can then be cast into an ordinary differntial
equation
\begin{align}\label{E1.22}
y''(z)-\frac{z}{2D}\,y'(z)+\frac{1}{2}\,u''(z)-\frac{z}{4D}\,u'(z)=0\;.
\end{align}
Eq.~(\ref{E1.22}) is the most general equation for $S_1$ with an
arbitrary scale-invariant potential $U(x,t)=\lambda U_1(x,t)$.
Once $y(z)$ is solved, the similarity solution $S_1(x,t)$ of
Eq.(\ref{E1.2}) is obtained by putting $z=x/\sqrt{t}$. To
illustrate the procedure, we shall solve the case of a class of
simple scale-invariant potentials.

\section{$U_1(x,t)=\mu\,x^p\,t^q$}

For definiteness we consider the simplest form that may have the
required scaling property, namely, $U_1(x,t)=\mu\,x^p\,t^q$ where
$\mu$, $p$ and $q$ are arbitrary real parameters.  Requiring that
$U_1$ be scale-invariant under the transformation (\ref{scale1})
with $b=2a$, we must have $q=-p/2$. From (\ref{E1.15}), we have
\begin{align}\label{E1.25}
u(z)=U_1(x,t)=\mu\,x^p\,t^q=\mu z^p\;.
\end{align}
Eq.(\ref{E1.22}) is then expressed as
\begin{align}\label{E1.27}
y''-\frac{z}{2D}\,y'=-\frac{\mu}{2}p(p-1)\,z^{p-2}+\frac{\mu\,p}{4D}\,z^{p}\;.
\end{align}

The general solution of Eq.(\ref{E1.27}) is
\begin{align}
y(z)=\alpha_0\,\int^z\,\exp\Big\{\frac{z'^2}{4D}\Big\}\,dz'+\alpha_1-\frac{\mu}{2}\,z^p
\;.
\end{align}
In order for the probability density $W(x,t)$ to be normalizable,
we require that $y(z)$ be finite as $z\to \pm\infty$.  As such,
$\alpha_0$ has to be set to zero, as the exponential term is
divergent in the domain of $z\in (-\infty,\infty)$.  Thus the
acceptable solution is
\begin{align}\label{E1.31}
y(z)=\alpha_1-\frac{\mu}{2}z^p\;.
\end{align}

Putting $z=x/\sqrt{t}$ back into (\ref{E1.31}) gives the solution
of the first order perturbation equation:
\begin{align}\label{E1.32}
S_1(x,t)=\alpha_1-\frac{\mu}{2}\Big(\frac{x}{\sqrt{t}}\Big)^p\;.
\end{align}

With $S_1(x,t)$ given in (\ref{E1.32}), the equation for
$S_2(x,t)$, i.e. Eq.(\ref{PA1.9}), is
\begin{align}\label{E1.33}
\dot{S}_2&=DS_2''-\frac{x}{t}S_2'\;.
\end{align}
By the procedure described in Sect.~4, this equation can be
reduced to
\begin{align}\label{E1.34}
y''-\frac{z}{2D}y'=0\;.
\end{align}
As discussed before, the finite solution is
$y(z)=\alpha_2=\mbox{constant}$. Hence $S_2$ is given by
$S_2(x,t)=\alpha_2$.

With $S_1$ and $S_2$ given above, the equations for $S_n$ ($n\geq
3$) turn out to have the same form as that for $S_2$, i.e.,
Eq.~(\ref{E1.33}). Therefore the solutions for $S_n$ ($n\geq 3$)
are $S_n(x,t)=\alpha_n=\mbox{constant}$.

From (\ref{PA1.11}), the general solution $W(x,t)$ to FPE in the
case $U_1=\mu x^pt^q$, where $q=-p/2$, is
\begin{align}\label{E1.35}
W(x,t)&\propto\mbox{exp}\Big\{-\frac{x^2}{4Dt}-\frac{\lambda\mu}
{D}\Big(\frac{x}{\sqrt{t}}\Big)^p+\frac{1}{D}\Big(\lambda
\alpha_1+\lambda^2 \alpha_2+\lambda^3
\alpha_3+\cdots\Big)\Big\}\notag\\
&\propto\mbox{exp}\Big\{-\frac{x^2}{4Dt}-\frac{\lambda\mu}{D}
\Big(\frac{x}{\sqrt{t}}\Big)^p\Big\}\;.
\end{align}
In the last step, the constants $\alpha_1$, $\alpha_2$, $\cdots$,
are set to zero since the part exp$\{\frac{1}{D}(\lambda
\alpha_1+\lambda^2 \alpha_2+\lambda^3 \alpha_3+\cdots)\}$ can be
absorbed into the normalization constant, which however can not be
determined until the parameters $p$ and $\mu$ are given.
Furthermore, for $W(x,t)$ to be normalizable, $p$ can only take
the values $p=0,1,2$, and $p={\rm even\  integer}$ for $p>2$.

Below we shall consider two special but interesting cases of such a form of drift potentials.

\subsection{$U_1=x/\sqrt{t}$, $p=1$ case}

Let us now consider the case with $\mu=1$ and $p=1$. The FPE is
\begin{align}\label{E1.36}
\frac{\partial W}{\partial t}=D\frac{\partial^2 W}{\partial
x^2}+\frac{\lambda} {\sqrt{t}}\frac{\partial W}{\partial x}\;.
\end{align}
Its similarity solution is
\begin{align}\label{E1.37}
W(x,t)=\frac{1}{\sqrt{4\pi Dt}}\,\mbox{exp}\Big\{-\frac{1}{4Dt}\Big(x+2\lambda\sqrt{t}
\Big)^2\Big\}\;.
\end{align}

This result is the same as that in [16]. One can see from
(\ref{E1.37}) that position $x$ is shifted as time $t$ elapses.

\subsection{$U_1=x^2/2t$, $p=2$ case}

The second example is the case with $U_1(x,t)=x^2/2t$, where the
parameters taken are $\mu=1/2$ and $p=2$. From the discussions
given before, we find that the probability distribution function
\begin{align}\label{E1.38}
W(x,t)=\sqrt{\frac{1+2\lambda}{4\pi Dt}}\,\mbox{exp}\Big\{-\frac{x^2}{4Dt}(1+2\lambda)\Big\}\;,
\end{align}
is the solution to the FPE
\begin{align}\label{E1.39}
\frac{\partial W}{\partial t}=D\frac{\partial^2 W}{\partial x^2}+\frac{\lambda x}{t}\frac{\partial W}
{\partial x}+\frac{\lambda}{t}W\;.
\end{align}
 In (\ref{E1.38}), one can see that the perturbation parameter
$\lambda$ controls the half width of the Gaussian distribution
$W(x,t)$.

\section{Summary}

When a partial differential equation possesses certain scaling
behavior, the so-called similarity method is of great help in
finding its solutions. One advantage of the similarity method is
to reduce the partial differential equation to an ordinary differential equation through
some new independent variables (called similarity variables),
which are certain combinations of the old independent variables.
The solutions so obtained, called similarity solutions of the
differential equations, likewise possess proper scaling forms. A
well-known example is the diffusion equation.

In this paper, we have applied the similarity method to a class of
perturbative FPE's with small time-dependent drift potentials
studied in [16].  We have presented the main ideas of the
similarity method.  The method was then applied to find similarity
solutions of the FPE with certain scale-invariant drift
potentials.  Our results show that similarity method can be a
useful tool to solve FPE with time-dependent drift and diffusion
coefficients.  While the present work is only concerned with the
perturbative FPE, it can be extended to the general FPE without
much difficulty \cite{LH2}.

\bigskip


\section*{Acknowledgments}

 This work is supported in part by the
National Science Council (NSC) of the Republic of China under
Grants NSC-99-2112-M-032-002-MY3 and  NSC-99-2811-M-032-012.

\end{document}